\documentclass[%
 reprint,
superscriptaddress,
preprintnumbers,
 amsmath,amssymb,
 aps,
]{revtex4-2}

\usepackage{subcaption}
\usepackage{graphicx}
\usepackage{dcolumn}
\usepackage{float}

\usepackage{bm}
\usepackage{booktabs}
\usepackage{siunitx}
\usepackage{hyperref}
\usepackage[mathlines]{lineno}

\begin{document}

\title{Observation of Non-Equilibrium Phonon Loss Across a Solid--Superfluid Interface}
\author{A.~Armatol} \affiliation{Univ. Lyon, Universit\'e Lyon 1, CNRS/IN2P3, IP2I-Lyon, F-69622, Villeurbanne, France}
\author{C. Augier} \affiliation{Univ. Lyon, Universit\'e Lyon 1, CNRS/IN2P3, IP2I-Lyon, F-69622, Villeurbanne, France}
\author{L.~Berg\'e}\affiliation{Universit\'e Paris-Saclay, CNRS/IN2P3, IJCLab, 91405 Orsay, France}
\author{J.~Billard} \affiliation{Univ. Lyon, Universit\'e Lyon 1, CNRS/IN2P3, IP2I-Lyon, F-69622, Villeurbanne, France}
\author{H.J.~Birch} \affiliation{Department of Physics, University of Zurich, Winterthurerstrasse  190, 8057 Zurich, Switzerland}
\author{J.~Bl\'e} \affiliation{Univ. Grenoble Alpes, CNRS, Grenoble INP, LPSC-IN2P3, 38000 Grenoble, France}
\author{C.L. Chang} \affiliation{Argonne National Laboratory, 9700 S Cass Ave, Lemont, IL 60439, USA} \affiliation{Kavli Institute for Cosmological Physics, The University of Chicago, 5640 S Ellis Ave., Chicago, IL 60637 USA} \affiliation{Department of Astronomy and Astrophysics, The University of Chicago, Eckhardt, 5640 S Ellis Ave., Chicago, IL 60637 USA}
\author{Y.-Y.~Chang} \affiliation{University of California Berkeley, Department of Physics, 366 LeConte Hall 7300, Berkeley, CA 94720, USA}
\author{L.~Chaplinsky} \affiliation{University of Massachusetts, Amherst Center for Fundamental Interactions and Department of Physics, 101 Stockbridge Hall, 80 Campus Center Way, Amherst, MA 01003-9337 USA}
\author{G.~Cline} \affiliation{Lawrence Berkeley National Laboratory, 1 Cyclotron Rd., Berkeley, CA 94720, USA}
\author{A.~Cochard} \affiliation{Universit\'e Paris-Saclay, CNRS/IN2P3, IJCLab, 91405 Orsay, France}
\author{I.~Cojocari} \affiliation{Universit\'e Paris-Saclay, CNRS/IN2P3, IJCLab, 91405 Orsay, France}
\author{J.~Colas} \affiliation{Univ. Lyon, Universit\'e Lyon 1, CNRS/IN2P3, IP2I-Lyon, F-69622, Villeurbanne, France}
\author{M.~De~Jesus} \affiliation{Univ. Lyon, Universit\'e Lyon 1, CNRS/IN2P3, IP2I-Lyon, F-69622, Villeurbanne, France}
\author{P.~de~Marcillac} \affiliation{Universit\'e Paris-Saclay, CNRS/IN2P3, IJCLab, 91405 Orsay, France}
\author{K.~Dwinger} \affiliation{Department of Physics, University of Zurich, Winterthurerstrasse  190, 8057 Zurich, Switzerland}
\author{R.~Faure} \affiliation{Univ. Lyon, Universit\'e Lyon 1, CNRS/IN2P3, IP2I-Lyon, F-69622, Villeurbanne, France}
\author{S.~Fiorucci} \affiliation{Lawrence Berkeley National Laboratory, 1 Cyclotron Rd., Berkeley, CA 94720, USA}
\author{M.~Garcia-Sciveres} \affiliation{Lawrence Berkeley National Laboratory, 1 Cyclotron Rd., Berkeley, CA 94720, USA} \affiliation{International Center for Quantum-field Measurement Systems for Studies of the Universe and Particles (QUP,WPI), High Energy Accelerator Research Organization (KEK), Oho 1-1, Tsukuba, Ibaraki 305-0801, Japan}
\author{J.~Gascon} \affiliation{Univ. Lyon, Universit\'e Lyon 1, CNRS/IN2P3, IP2I-Lyon, F-69622, Villeurbanne, France}
\author{C.~Girard-Carillo} \affiliation{Univ. Grenoble Alpes, CNRS, Grenoble INP, LPSC-IN2P3, 38000 Grenoble, France}
\author{W.~Guo} \affiliation{Department of Mechanical Engineering, FAMU-FSU College of Engineering, Florida State University, 2525 Pottsdamer Street,  Tallahassee, FL 32310, USA} \affiliation{National High Magnetic Field Laboratory, 1800 E Paul Dirac Dr., Tallahassee, FL 32310, USA}
\author{L.~Haegel} \affiliation{Univ. Lyon, Universit\'e Lyon 1, CNRS/IN2P3, IP2I-Lyon, F-69622, Villeurbanne, France}
\author{S.J.~Haselschwardt} \affiliation{University of Michigan, Randall Laboratory of Physics, Ann Arbor, MI 48109-1040, USA}
\author{S.A.~Hertel} \affiliation{University of Massachusetts, Amherst Center for Fundamental Interactions and Department of Physics, 101 Stockbridge Hall, 80 Campus Center Way, Amherst, MA 01003-9337 USA}
\author{K.~Hunter} \affiliation{Texas A\&M University, Department of Physics and Astronomy, 4242 TAMU, 578 University Dr., College Station, TX 77843-4242, USA}
\author{L.~Juigne} \affiliation{Department of Physics, University of Zurich, Winterthurerstrasse  190, 8057 Zurich, Switzerland}
\author{A.~Juillard} \affiliation{Univ. Lyon, Universit\'e Lyon 1, CNRS/IN2P3, IP2I-Lyon, F-69622, Villeurbanne, France}
\author{A.~Kavner} \affiliation{Department of Physics, University of Zurich, Winterthurerstrasse  190, 8057 Zurich, Switzerland}
\author{J.~Lamblin} \affiliation{Univ. Grenoble Alpes, CNRS, Grenoble INP, LPSC-IN2P3, 38000 Grenoble, France}
\author{T.~Le-Bellec} \affiliation{Univ. Lyon, Universit\'e Lyon 1, CNRS/IN2P3, IP2I-Lyon, F-69622, Villeurbanne, France}
\author{X.~Li} \affiliation{Lawrence Berkeley National Laboratory, 1 Cyclotron Rd., Berkeley, CA 94720, USA}
\author{J.~Lin} \affiliation{University of California Berkeley, Department of Physics, 366 LeConte Hall 7300, Berkeley, CA 94720, USA} \affiliation{Lawrence Berkeley National Laboratory, 1 Cyclotron Rd., Berkeley, CA 94720, USA}
\author{R.~Mahapatra} \affiliation{Texas A\&M University, Department of Physics and Astronomy, 4242 TAMU, 578 University Dr., College Station, TX 77843-4242, USA}
\author{S.~Marnieros} \affiliation{Universit\'e Paris-Saclay, CNRS/IN2P3, IJCLab, 91405 Orsay, France}
\author{C.A.~Marrache-Kikuchi} \affiliation{Universit\'e Paris-Saclay, CNRS/IN2P3, IJCLab, 91405 Orsay, France}
\author{N.~Martini} \affiliation{Univ. Lyon, Universit\'e Lyon 1, CNRS/IN2P3, IP2I-Lyon, F-69622, Villeurbanne, France}
\author{W.~Matava} \email{wmatava@berkeley.edu} \affiliation{University of California Berkeley, Department of Physics, 366 LeConte Hall 7300, Berkeley, CA 94720, USA} \affiliation{Lawrence Berkeley National Laboratory, 1 Cyclotron Rd., Berkeley, CA 94720, USA}
\author{D.N.~McKinsey} \affiliation{University of California Berkeley, Department of Physics, 366 LeConte Hall 7300, Berkeley, CA 94720, USA} \affiliation{Lawrence Berkeley National Laboratory, 1 Cyclotron Rd., Berkeley, CA 94720, USA}
\author{J.~Menu} \affiliation{Univ. Grenoble Alpes, CNRS, Grenoble INP, LPSC-IN2P3, 38000 Grenoble, France}
\author{K.~Moraa} \affiliation{Department of Physics, University of Zurich, Winterthurerstrasse  190, 8057 Zurich, Switzerland}
\author{V.~Novati} \affiliation{Univ. Grenoble Alpes, CNRS, Grenoble INP, LPSC-IN2P3, 38000 Grenoble, France}
\author{E.~Olivieri}\affiliation{Universit\'e Paris-Saclay, CNRS/IN2P3, IJCLab, 91405 Orsay, France}
\author{B.~Penning} \affiliation{Department of Physics, University of Zurich, Winterthurerstrasse  190, 8057 Zurich, Switzerland}
\author{H.~Pinckney} \affiliation{University of Massachusetts, Amherst Center for Fundamental Interactions and Department of Physics, 101 Stockbridge Hall, 80 Campus Center Way, Amherst, MA 01003-9337 USA}
\author{M.~Platt} \affiliation{Texas A\&M University, Department of Physics and Astronomy, 4242 TAMU, 578 University Dr., College Station, TX 77843-4242, USA}
\author{M.~Pyle} \affiliation{University of California Berkeley, Department of Physics, 366 LeConte Hall 7300, Berkeley, CA 94720, USA} \affiliation{Lawrence Berkeley National Laboratory, 1 Cyclotron Rd., Berkeley, CA 94720, USA}
\author{D.~Poda}\affiliation{Universit\'e Paris-Saclay, CNRS/IN2P3, IJCLab, 91405 Orsay, France}
\author{Y.~Qi} \affiliation{Department of Mechanical Engineering, FAMU-FSU College of Engineering, Florida State University, 2525 Pottsdamer Street,  Tallahassee, FL 32310, USA} \affiliation{National High Magnetic Field Laboratory, 1800 E Paul Dirac Dr., Tallahassee, FL 32310, USA}
\author{M.~Reed} \affiliation{University of California Berkeley, Department of Physics, 366 LeConte Hall 7300, Berkeley, CA 94720, USA}
\author{R.K.~Romani} \affiliation{University of California Berkeley, Department of Physics, 366 LeConte Hall 7300, Berkeley, CA 94720, USA} 
\author{I.~Rydstrom} \affiliation{University of California Berkeley, Department of Physics, 366 LeConte Hall 7300, Berkeley, CA 94720, USA}
\author{B.~Sadoulet} \affiliation{University of California Berkeley, Department of Physics, 366 LeConte Hall 7300, Berkeley, CA 94720, USA}
\author{S.~Scorza} \affiliation{Univ. Grenoble Alpes, CNRS, Grenoble INP, LPSC-IN2P3, 38000 Grenoble, France}
\author{B.~Serfass} \affiliation{University of California Berkeley, Department of Physics, 366 LeConte Hall 7300, Berkeley, CA 94720, USA}
\author{P.~Sorensen} \affiliation{Lawrence Berkeley National Laboratory, 1 Cyclotron Rd., Berkeley, CA 94720, USA}
\author{S.~Steinfeld} \affiliation{University of California Berkeley, Department of Physics, 366 LeConte Hall 7300, Berkeley, CA 94720, USA} \affiliation{Lawrence Berkeley National Laboratory, 1 Cyclotron Rd., Berkeley, CA 94720, USA}
\author{H.~Su} \affiliation{University of Massachusetts, Amherst Center for Fundamental Interactions and Department of Physics, 101 Stockbridge Hall, 80 Campus Center Way, Amherst, MA 01003-9337 USA}
\author{A.~Suzuki} \affiliation{Lawrence Berkeley National Laboratory, 1 Cyclotron Rd., Berkeley, CA 94720, USA}
\author{R.L.~Vaughn~II} \affiliation{University of Massachusetts, Amherst Center for Fundamental Interactions and Department of Physics, 101 Stockbridge Hall, 80 Campus Center Way, Amherst, MA 01003-9337 USA}
\author{C.~Veihmeyer} \affiliation{University of California Berkeley, Department of Physics, 366 LeConte Hall 7300, Berkeley, CA 94720, USA}
\author{V.~Velan} \affiliation{Lawrence Berkeley National Laboratory, 1 Cyclotron Rd., Berkeley, CA 94720, USA}
\author{G. Wang} \affiliation{Argonne National Laboratory, 9700 S Cass Ave, Lemont, IL 60439, USA}
\author{P.~Vittaz} \affiliation{Univ. Lyon, Universit\'e Lyon 1, CNRS/IN2P3, IP2I-Lyon, F-69622, Villeurbanne, France}
\author{Y.~Wang} \affiliation{University of California Berkeley, Department of Physics, 366 LeConte Hall 7300, Berkeley, CA 94720, USA} \affiliation{Lawrence Berkeley National Laboratory, 1 Cyclotron Rd., Berkeley, CA 94720, USA}
\author{M.R. Williams} \email{michaelwilliams@lbl.gov} \affiliation{Lawrence Berkeley National Laboratory, 1 Cyclotron Rd., Berkeley, CA 94720, USA}
\author{J.~Wuko} \affiliation{University of Massachusetts, Amherst Center for Fundamental Interactions and Department of Physics, 101 Stockbridge Hall, 80 Campus Center Way, Amherst, MA 01003-9337 USA}

\collaboration{TESSERACT Collaboration}

\date{\today}

\begin{abstract}

We measure the performance of an athermal phonon calorimeter in vacuum and in thermal contact with superfluid $^4$He. Using a cryogenic photon calibration system, we show that the superfluid suppresses the collection of the phonon signal by over 99\%.  We conclude that submerged sensors will not benefit TESSERACT's superfluid helium dark matter target, but that superfluid helium may be useful to superconducting qubits as a sink for athermal phonons generated by radiation and other spontaneous sources. Our measurement is the first probe of solid-to-superfluid phonon transmission at energies higher than those stably supportable by the superfluid. We discuss thermal flow inside the  calorimeter, as well as systematics from an alternative calibration scheme.

\end{abstract}


\maketitle

\section{Introduction} \label{sec:introduction}

The search for low-mass dark matter induced nuclear recoils in terrestrial experiments requires the development of new technologies to overcome poor kinematic matching and low recoil energies when compared with more massive dark matter models such as the WIMP \cite{Bertone2005}. Recently, superfluid $^4$He has gained interest as a potential dark matter recoil target due to its low nuclear mass, interesting signal channels, ease of chemical purification, and intrinsic scalability~\cite{Hertel_He_DM,DELight_conceptual,HeRALD_Pinckney,Schutz_2016}. We have operated a detector that uses un-wetted athermal phonon calorimeters to detect recoils in a superfluid $^4$He bath sitting below~\cite{HeRALD_Pinckney}. An additional array of submerged sensors would improve signal reconstruction by enhancing collection of prompt scintillation photons or triplet excimer decays \cite{Carter_triplet_det}. 

We have developed calorimeters with energy resolutions as low as 253 meV \cite{WilliamsFastNeutron,Bui_2025}. The extreme sensitivity of these low-threshold athermal phonon detectors has provided a window into non-equilibrium energy transport in solids, which is also an important consideration for superconducting quantum circuits~\cite{AnthonyPetersen2024,ddyy-t37l}. In superconducting qubits, the same non-equilibrium processes contribute to excess quasiparticle densities in superconducting films, leading to premature decoherence of the qubit state. Band-gap engineering \cite{Sun_2012,McEwen_2024,Marchegiani_2022} and large-area normal metal films \cite{Iaia2022PhononDownconversion} have been successfully deployed to reduce the impact of non-equilibrium energy in superconducting qubit circuits; however, further improvements are still required. The unique properties of superfluid helium motivate exploring its potential as an additional channel for removing non-equilibrium energy from superconducting circuits. Understanding energy transport across the solid–helium interface is therefore essential.

Key to evaluating this possibility, and to the performance of submerged calorimeters, is the behavior of athermal phonons at the solid–helium interface. Phonon transmission at solid--superfluid interfaces has long been of interest to the condensed matter community due to anomalously high transmission probabilities despite severe acoustic mismatch \cite{Swartz_Pohl}. Historically, interest has gone towards the study of conduction via the stable quasiparticles present in the superfluid below 1.49 meV (phonons and rotons) \cite{Amrit_resonance, Kapitza1941,Kapitza_tension, Adamenko_roughness_resonance, Bandler_thesis,Wyatt_NaF_transmission, Wyatt_NaF_transmission_freqdep}. Notably, solid materials can support phonons with energies of $\mathcal{O}(100)$ meV. Until now, no observation of the transmission of athermal phonons with energies greater than $1.49$~meV has been made.

In this Letter, we describe the operation and performance of a TESSERACT athermal phonon sensor, hereafter referred to as a pixel phonon detector (PPD), when in contact with superfluid $^4$He.  We investigate both the flow of thermal and athermal phonons in and out of the PPD. We demonstrate an efficient transmission of athermal phonons across the solid--superfluid interface, with the sensor losing almost all of the energy it absorbs. We discuss our measurements' relationships to both theory and previous measurements, their impact on the development of future low-mass dark matter experiments, and superfluid $^4$He's utility as a phonon sink for superconducting qubits.
\section{Experimental Setup} \label{sec:hardware}
\begin{figure}
    \centering
    \includegraphics[width=1.0\linewidth]{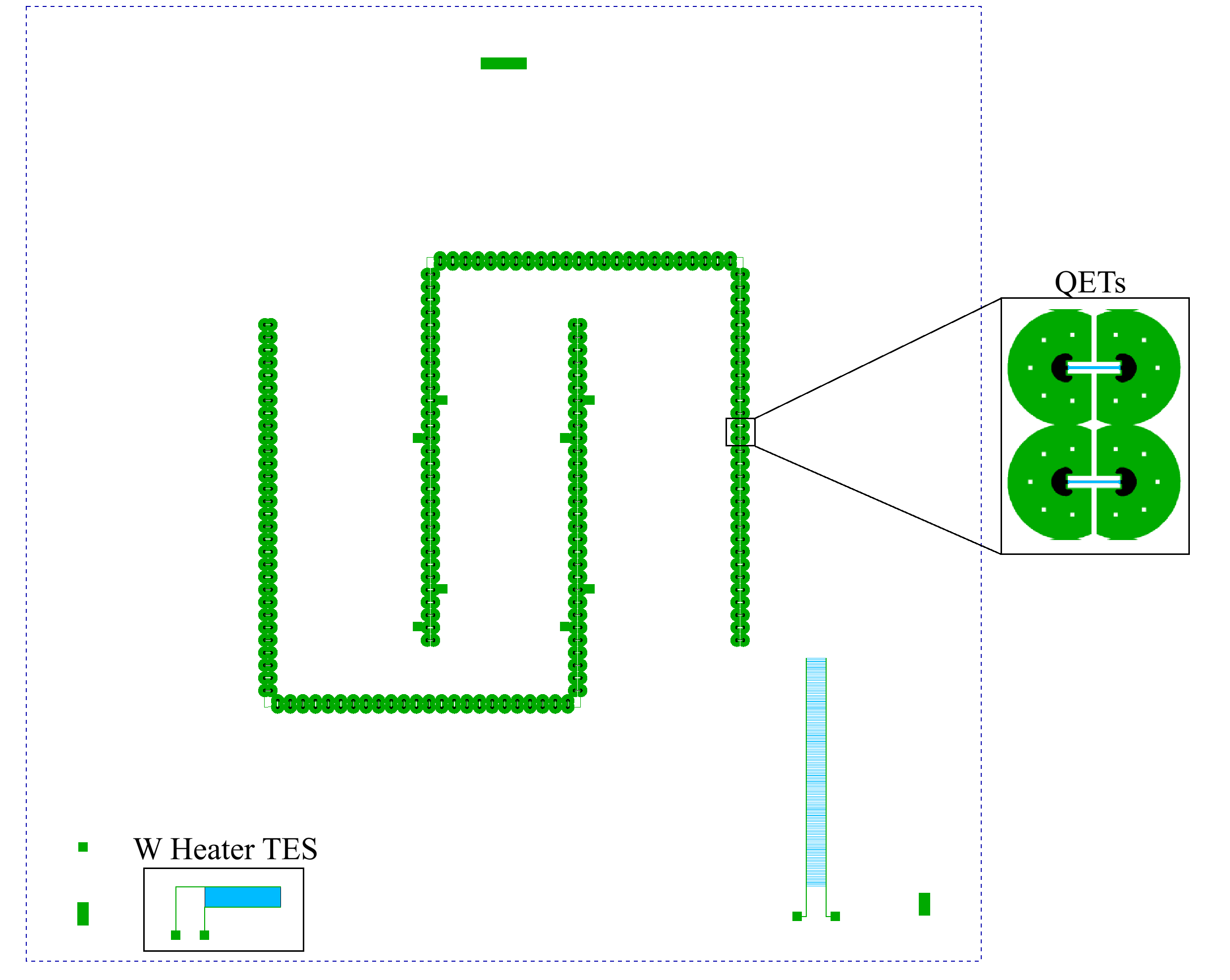}
    \caption{Layout of the PPD. Green areas denote aluminum, cyan denotes tungsten, and black areas denote areas of tungsten and aluminum overlap. White space is bare silicon, and the dashed line denotes the borders of the cm$^2$ silicon chip. The two U-shaped QET chains were connected with aluminum wirebonds (not shown) and read out as a single channel. The TES in the bottom left corner was also read out and biased to heat the PPD when needed (see Appendix~\ref{appendix:Gab}). The TES in the bottom right was not instrumented.}
    \label{fig:PPD_diagram}
\end{figure}

The PPD is composed of voltage-biased tungsten Transition Edge Sensors (TESs)~\cite{irwinTransitionEdgeSensors2005} coupled to aluminum phonon collection fins in the Quasiparticle-trap-assisted Electrothermal-feedback Transition-edge-sensor (QET)~\cite{irwinQuasiparticleTrapAssisted1995} architecture, fabricated on a 1 cm$^2$ by 1 mm thick intrinsic silicon substrate. We wire 172 QETs into a single readout channel, giving a total sensitive surface area coverage of 1.5\%. On the same surface are two separate tungsten TESs of 200 µm by 800 µm by 60 nm, one of which is used as a heater for the thermal studies discussed in Appendix~\ref{appendix:Gab}, and will be referred to as a heater. A diagram of the PPD can be seen in Fig.~\ref{fig:PPD_diagram}. Both the QET and heater channels are read out by two-stage DC SQUID amplifiers.

The PPD is attached to a gold-plated copper mounting structure via rubber cement at three corners, with aluminum wirebonds providing electrical connection to a readout PCB. A 25.4 µm diameter gold wirebond links an aluminum pad on the PPD to its mounting structure, providing a thermal link to the bath. The mounting structure is housed in a superfluid $^4$He-tight experimental detector cell. A calibrated ruthenium oxide thermometer is bolted to the cell adjacent to the mounting structure. A single-mode optical fiber is coupled to the cell via a superfluid-tight feedthrough, providing an in-situ method of calibrating the detector with $3.06$~eV photons generated by a laser diode at room temperature. The detector platform is mounted below the mixing chamber plate of a dilution refrigerator, with a base temperature of $\sim7$~mK. $99.999\%$ pure helium of natural isotopic abundance \cite{Isotopic_abundances} was introduced into the cell through a 0.89 mm inner diameter stainless steel capillary thermalized to the cryostat's various temperature stages. Fill heights were calculated from the pressure drop in storage cylinders and the geometry of the cell. 

The PPD was first characterized while dry, after which a small amount of $^4$He was introduced so that only a film of superfluid coated the PPD. The sensor was then re-characterized in this \textit{film} state. Finally, sufficient $^4$He was introduced to fully submerge the PPD, and the process was repeated for this \textit{bulk} state.

\section{PPD Characterization}

\subsection{TES Electrothermal Preliminaries} \label{sec:methods_IV_dIdV}

\begin{figure}
    \centering
    \includegraphics[width=\linewidth]{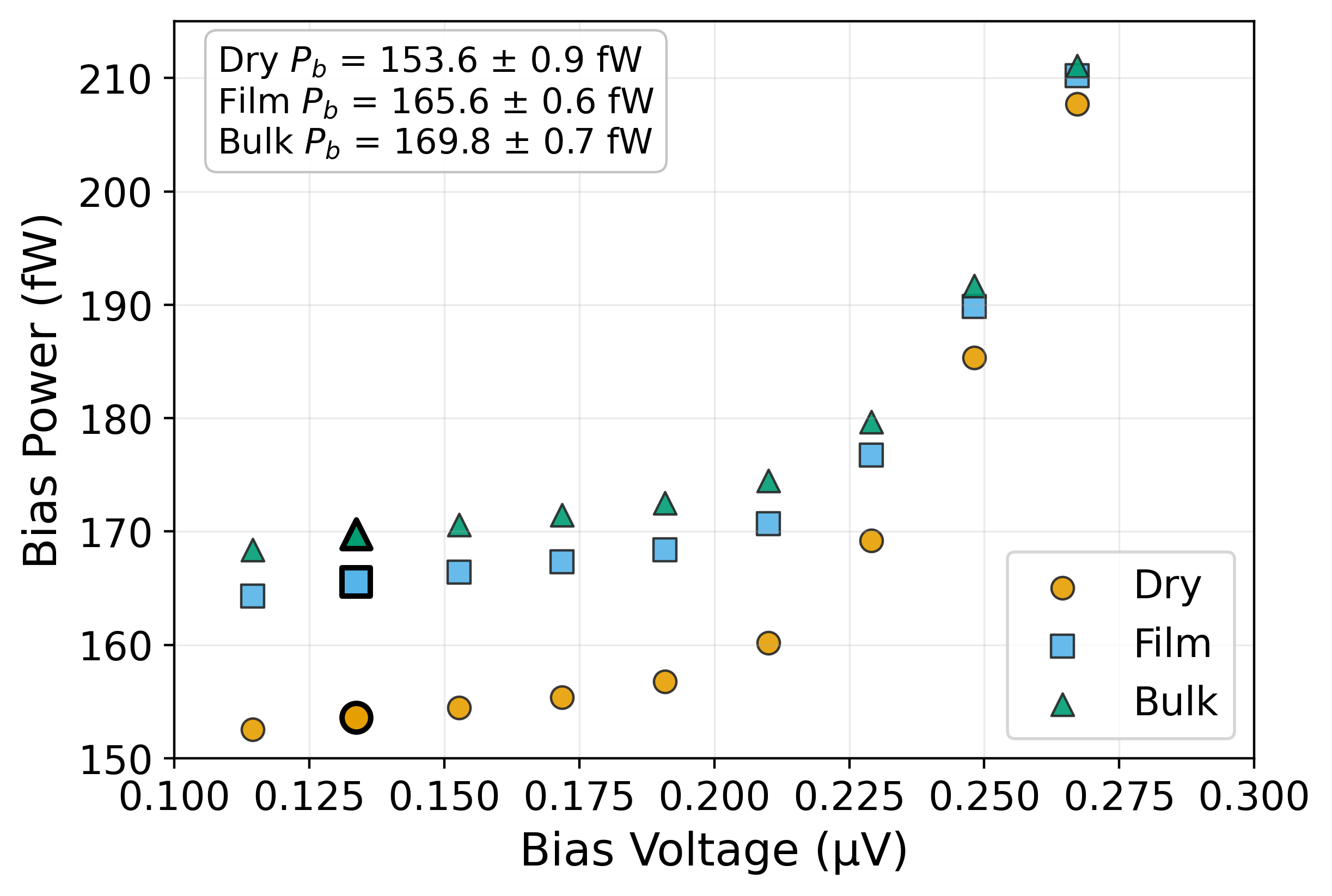}
    \caption{TES bias power vs. bias voltage. Dry data are represented by orange circles, film data by blue squares, and bulk data by green triangles. The bias point for transition operation described in Sec.~\ref{sec:methods_athermal} bears a thicker boundary line. Bias powers $P_b$ at this operation point are recorded in the inset.}
    \label{fig:IV}
\end{figure}

\begin{figure}
    \centering
    \includegraphics[width=1.0\linewidth]{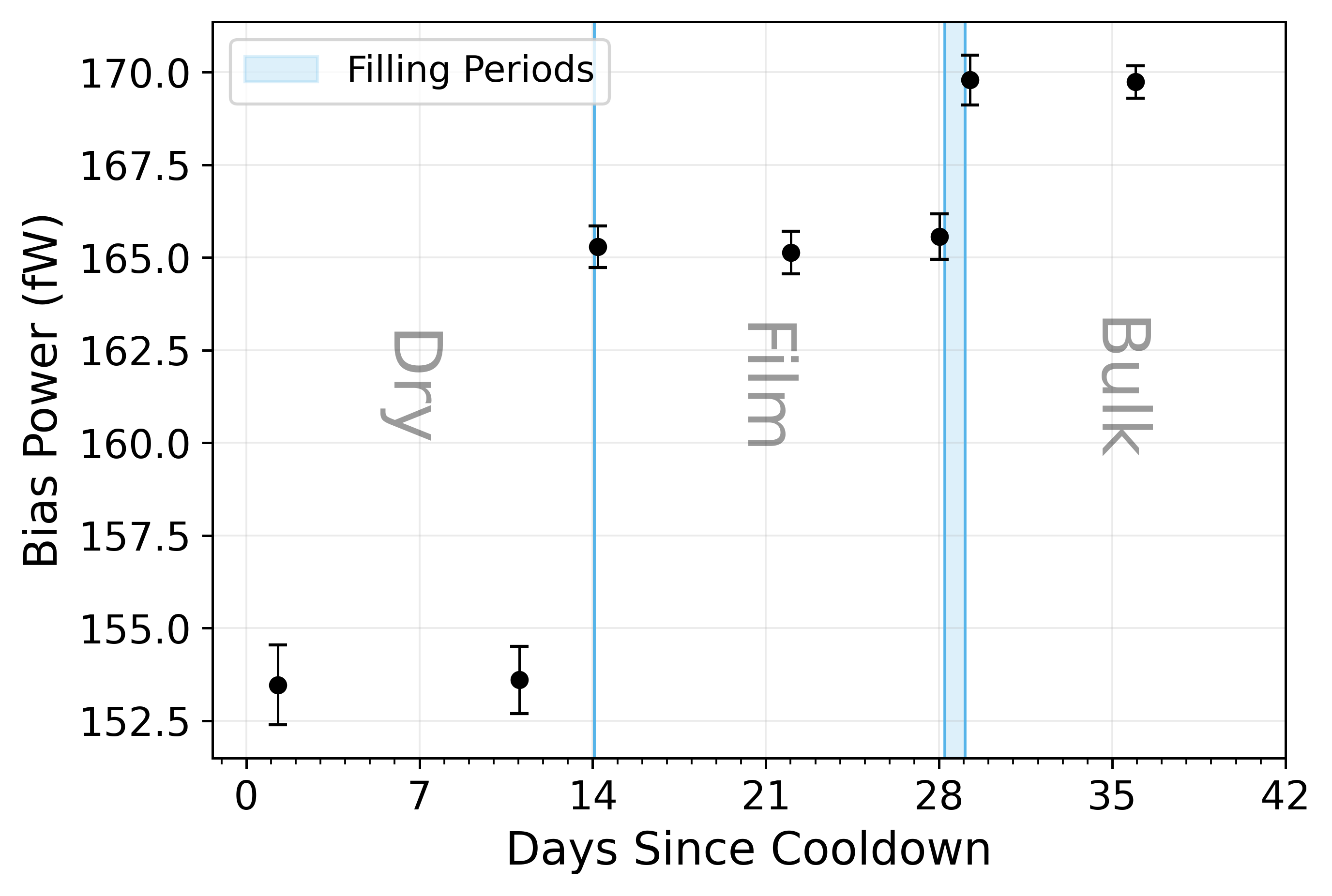}
    \caption{Measured bias power in transition with constant voltage bias throughout the run. Periods of helium filling are shaded in blue. Error bars represent statistical uncertainties. }
    \label{fig:bias powers}
\end{figure}

In each of the three fill states, we performed I-V scans, sweeping downward in bias voltage. From the measured I-V curves, we extract the circuit parasitic resistance ($R_p$) and the TES resistance ($R_0$) and bias power ($P_0$) at each point along the curve. At a bias voltage in transition fixed across the fill states, we injected a square wave over the DC bias and calculated the circuit's complex admittance ($\frac{dI}{dV}(\omega)$) based from the response. Both $\text{Re}(\frac{dI}{dV}(\omega))$ and $\text{Im}(\frac{dI}{dV}(\omega))$ were then simultaneously fit to an analytical model of the TES's complex admittance~\cite{watkinsThesis}. The sensor's current-to-power transfer function ($\frac{dP}{dI}(\omega))$ was then calculated~\cite{irwinTransitionEdgeSensors2005, watkinsThesis}. The PPD's bias power as a function of bias voltage can be seen in Fig.~\ref{fig:IV}. The device's bias power in transition has a measured increase of 12 fW in the film state, with an additional 4 fW gained in the bulk state. These measurements were carried out multiple times at each fill state. In Fig.~\ref{fig:bias powers} we plot a larger set of bias power measurements throughout the run vs.\ time.

\subsection{Characterization of Athermal Phonon Collection} \label{sec:methods_athermal}

The PPD was biased in its transition at a fixed bias voltage across the three fill states. To calibrate the PPD response to athermal phonons in the dry state we injected $1$ µs-wide pulses of $3.06$~eV photons into the experimental cell and onto the PPD's reverse (bare) face. The laser power was tuned so that $\mathcal{O}(1)$ photon was absorbed by the silicon in each pulse. The PPD's response to injected photons was measured offline using an optimal (matched) filter (OF) \cite{OF_theory}, with timing determined by a synchronization signal from the function generator driving the laser.

\begin{figure}
    \centering
    \includegraphics[width=1.0\linewidth]{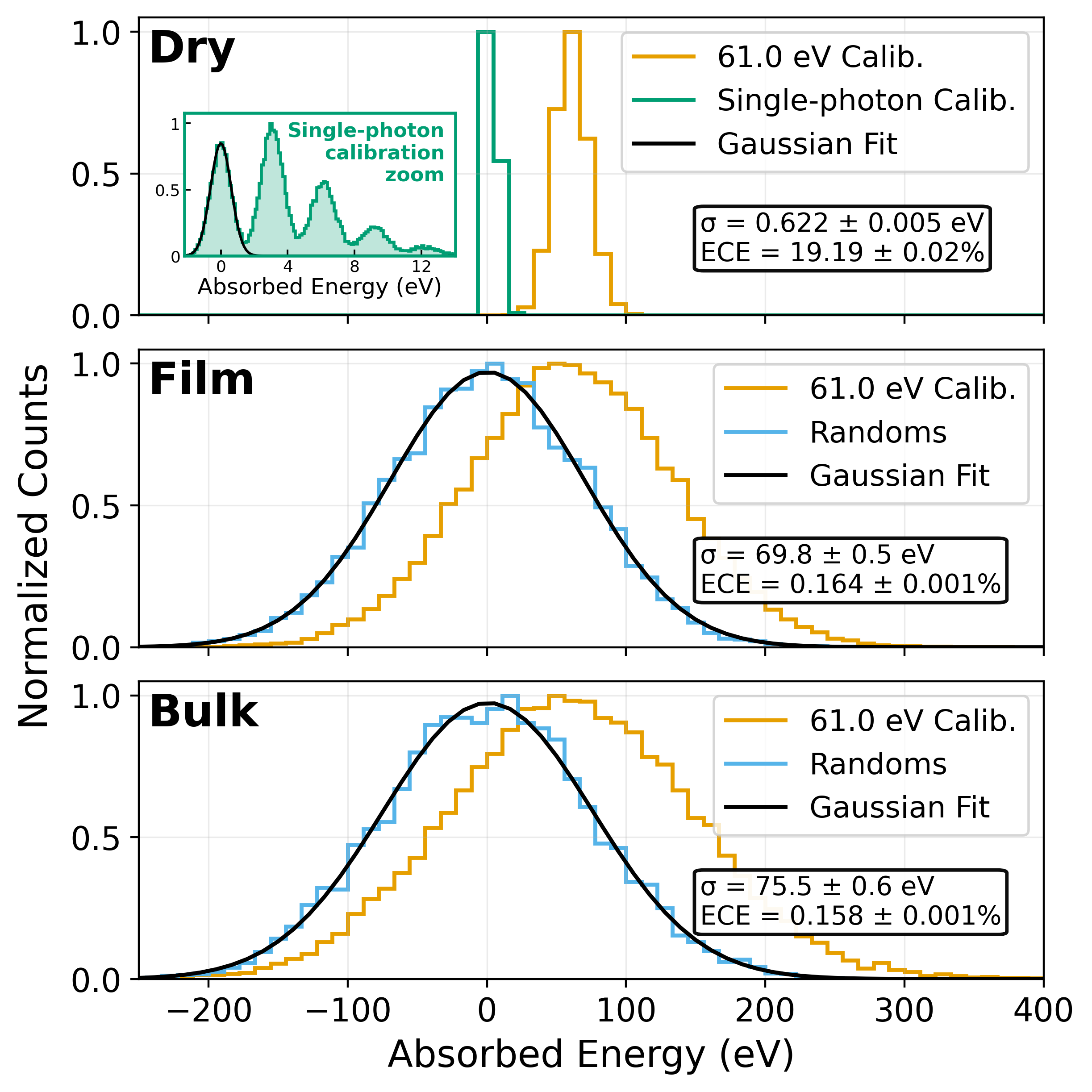}
    \caption{Calorimeter response in three physical states: dry, coated with superfluid film, and immersed in bulk superfluid. Responses to pulses of $3.06$~eV photons with a mean of $61.0$~eV, applied in each state, are shown in orange. The single photon calibration distribution is also shown for the dry calorimeter in green. Finally, the distribution of random events with the laser off is plotted in blue for the two wet datasets, with Gaussian fits superimposed, along with a Gaussian fit to the dry 0-photon peak. The corresponding distribution is not shown for the dry state; however, it is equivalent to the 0-photon peak seen in the inset.}
    \label{fig:wet calibration}
\end{figure}

The detector, when dry, exhibits a sub-photon energy resolution of $0.622$~eV, enabling a direct calibration to the photon energy absorbed into the silicon. The distribution of measured PPD responses is seen in the inset of Fig.~\ref{fig:wet calibration}. The measurement of $\frac{dP}{dI}(\omega)$ allows for the measurement of the PPD's \textit{energy collection efficiency} (ECE), the fraction of the energy deposited in the silicon that is ultimately collected by the TES. In the dry state, the measured ECE is 19.2\%, in line with previous measurements of similar PPDs given additional losses into the mounting glue and additional inert metal~\cite{WilliamsFastNeutron}.

Photon calibration data when dry were also taken at a higher laser intensity, at which the mean number of photons absorbed by the detector per laser pulse was much higher. By virtue of linearity of the PPD in this energy range, the average number of absorbed photons in this higher intensity dataset could be determined using the aforementioned calibration. Sensor linearity was confirmed by stability of OF's $\chi^2$ goodness-of-fit metric over the range of amplitudes. The distribution of responses is shown in the top panel of Fig.~\ref{fig:wet calibration}, corresponding to a mean absorbed energy of $61.0$~eV ($\sim$ 20 photons).

In both of the wet states, we reproduced this higher intensity dataset with identical laser settings. Under the assumption that the helium does not significantly perturb the average number of absorbed photons, the mean response is assumed to correspond to $61.0$~eV of absorbed energy. The response distributions are shown in the two lower panels of Fig.~\ref{fig:wet calibration}. From this calibration and the measurements of $\frac{dP}{dI}(\omega)$, we compute ECEs of 0.164$\%$ and 0.158$\%$ for the film and bulk states respectively.

In each of the two wet states, data with the laser off were collected. At randomly triggered times, the best-fit pulse amplitude was calculated with the same OF as with the $61.0$~eV calibration data, with the distributions shown for the two wet states in the lower two panels of Fig.~\ref{fig:wet calibration}. Defining the resolution as the standard deviation of this distribution, we measured resolutions of $(69.8 \pm 0.5)$ and $(75.5\pm 0.6)$~eV for the film and bulk cases, considerably poorer than the dry analog of $(0.622\pm0.005)$~eV.

For all three fill states, the mean PPD pulse shape from incident $3.06$~eV photons was measured. These observed pulse shapes were then convolved with $\frac{dP}{dI}(\omega)$ to yield the time distribution of energy arrivals into the tungsten TES. Finally, the theoretical impulse-response function for the TES was calculated from $\frac{dP}{dI}(\omega)$. These three characteristic pulse shapes are summarized in Fig.~\ref{fig:pulse_shapes}. When dry, the PPD responds to photon absorptions with best-fit fall times $\tau_1=(43.5\pm0.2)$ and $\tau_2=(364\pm2)$~µs. In the film state, the fall times are $\tau_1=(30.7\pm0.1)$ and $\tau_2=(302\pm1)$~µs. In the bulk state, the fall times are $\tau_1 = (28.0\pm0.1)$ and $\tau_2 = (264\pm2)$~µs. Notably, the TES impulse response pulse shape is a weak function of the fill state, while energy arrives to the tungsten TES much faster on average when wet.

\section{Discussion} \label{sec:discussion}

\subsection{Helium's Impact on TES Bias Power} \label{sec:discussion_IV}

Section~\ref{sec:methods_IV_dIdV} demonstrates that the PPD's bias power increases substantially in response to a thin superfluid film, with further modest improvements seen when fully submerged. Figure~\ref{fig:bias powers} also demonstrates that the change is due to the helium, and not a gradual trend in bias power over the course of the run. There are two possibilities that explain this observation. In the first, the superfluid cools the tungsten TES via conduction. In the second, the cooling stems from a reduction in parasitic power. Collection of athermal phonons generated by the ``low energy excess'' (LEE) \cite{LEEReview} has been demonstrated as a source of parasitic power~\cite{TwoChannelPaper, Run57Paper}. Since we have shown that the superfluid significantly hinders phonon collection, a corresponding reduction in this parasitic power is expected and may explain all of the measured change in bias power. For further discussion on this topic, and discussion of the thermal conductance between the helium, PPD metal films, and PPD substrate, see Appendix~\ref{appendix:Gta}.

\subsection{Helium's Impact on Phonon Collection} \label{sec:discussion_athermal}

\begin{figure}
    \centering
    \includegraphics[width=1.0\linewidth]{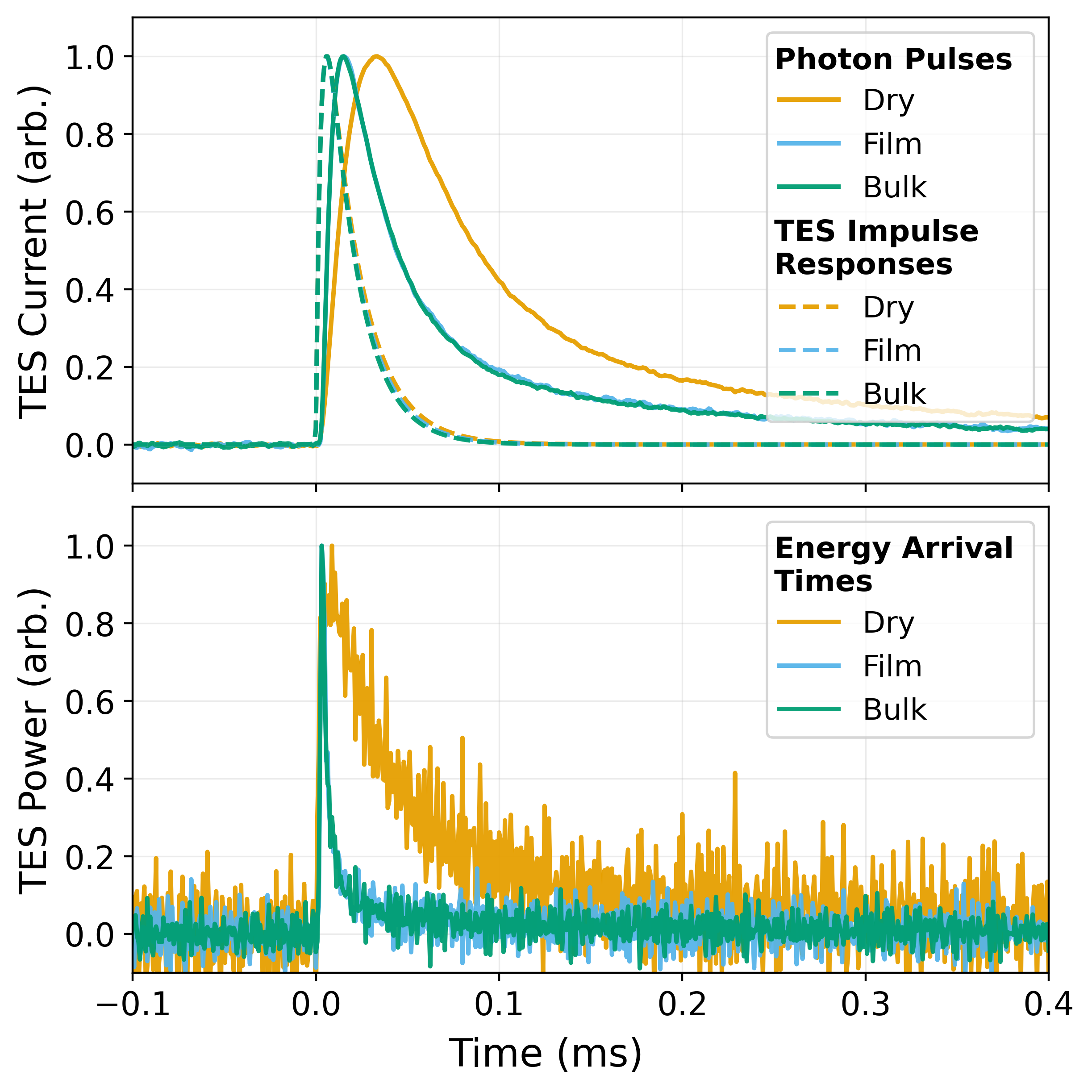}
    \caption{Characteristic pulse shapes. (Top) Average measured response to photon absorptions (solid) and TES impulse responses predicted from $\frac{dP}{dI}(\omega)$ (dashed). (Bottom) Energy arrival times in the tungsten TES, generated by convolving the average measured pulse shape templates with $\frac{dP}{dI}(\omega)$.}
    \label{fig:pulse_shapes}
\end{figure}

Section~\ref{sec:methods_athermal} demonstrates that our PPD's ECE degrades by over two orders of magnitude when coated or immersed in superfluid helium. Since superfluid has been demonstrated not to perturb quasiparticle diffusion in aluminum films~\cite{Osterman2023}, and since the TES's impulse response is largely unperturbed by the helium, efficient transmission of athermal phonons into the superfluid is the only credible loss mechanism. This interpretation is further supported by the change in pulse shape demonstrated in Fig.~\ref{fig:pulse_shapes}: the faster energy arrival to the tungsten TES when wet are consistent with preferential collection of phonons that take a near straight-line path to an aluminum film, with those that are not initially incident upon aluminum films escaping. 

We note that the observed signal losses are too large to be explained by standard phonon-roton transmission as described in \cite{Tanatarov_theory}. First, consider as a toy model a dry PPD without loss mechanisms (e.g. sub-gap downconversion, phonon escape, pair breaking in other metals, quasiparticle recombination, etc.). In this unrealistic limit, the ECE is 100\%. When the same toy PPD is submerged into a superfluid  bath we have added a potential loss mechanism. Assume a probability of 1\% for athermal phonon transmission from solid surfaces into the helium~\cite{Tanatarov_theory}. Athermal phonons will be assumed to have a probability of 40\% of transmission at silicon/aluminum interfaces \cite{Martinez_Si_Al_transmission}, and probability 70\% of breaking a Cooper pair on a two-way trip through the aluminum \cite{Moore_thesis_Al_pairbreakinglength}. Based on the above probabilities, an athermal phonon incident on an aluminum film has a 34\% chance of breaking a Cooper pair. Likewise, it has a probability of 0.3\% of escaping into the helium after traversing the aluminum, and 66\% of returning to the silicon. (We calculate these using the above assumptions and an arbitrary number of bounces at the Si/Al and Al/He interfaces). A given bounce (not necessarily upon the aluminum) therefore has a probability 0.5\% of breaking Cooper pairs and 1.0\% of escaping, accounting for both transmission probabilities and aluminum coverage fraction. We therefore expect an ECE of order 30\%, degraded by only a factor of $\sim 3$ from the dry toy model.

While crude, this toy model is conservative: the complex loss mechanisms it ignores either scale with phonon lifetime/number of bounces and preferentially degrade the collection of the dry case or affect the two cases roughly identically. Our estimate of a 3$\times$ reduction in collection efficiency due to helium is therefore a worst-case scenario, and in substantial tension with our observations.

In the above toy model we assumed that the loss occurs through ballistic phonons that have consecutive bounces at random positions on the PPD's surface. For a silicon PPD of our size, this corresponds to energies of $5$~meV and below \cite{SiBallCutoff}. The baseline estimate of $\mathcal{O}(1)\%$ transmission into the helium derived in \cite{Tanatarov_theory} is calculated only up to 1.49 meV, above which the superfluid does not support stable excitations. If the loss indeed occurs from ballistic phonons, our results suggest that phonons with energies below 5 meV, including a substantial population greater than 1.49 meV~\cite{Kim_Si_Phonon_DOS_2015}, have transmission probabilities significantly greater than 1\%. If instead the loss occurs from diffusive phonons, our results instead suggest a non-negligible probability of transmission for phonons with energy greater than 5 meV, although the probability in this case is more difficult to bound. In either case, we expect the loss to manifest primarily through phonons with energies greater than 1.49 meV. While anomalously high transmission at superfluid--solid interfaces has been observed at both phonon \cite{Wyatt_NaF_transmission,Wyatt_NaF_transmission_freqdep} and roton \cite{Bandler_thesis} scales, to our knowledge, no experimental or theoretical studies have probed energies greater than 1.49 meV.

\subsection{Ramifications}

\subsubsection{Helium-Target Dark Matter Searches}

The conclusions drawn regarding the resolution of wet PPDs have large implications for future dark matter searches using superfluid helium targets \cite{Schutz_2016,DELight_conceptual, Hertel_He_DM, HeRALD_Pinckney}. Experiments that use a submerged array of detectors to sense downward-traveling $\sim 16$~eV photons must either develop sensors less prone to phonon leakage, or pivot to relying on a non-wetted top array only \cite{DELight_conceptual}.  In the latter case, the resulting loss of sensor coverage will lower the collection efficiency of quanta produced in the helium, resulting in increased detector threshold and weakened sensitivity to low-mass dark matter. TESSERACT will develop a low-background superfluid helium detector that will operate without submerged PPDs, while exploring paths to overcome phonon leakage into the helium for next-generation experiments.

\subsubsection{Superconducting Qubits}

 The realization of quantum computers based on superconducting quantum circuits is currently dependent on the ability to correct errors impacting individual qubits~\cite{Fowler_2012}. Such ``fault tolerant'' systems have been deployed to various levels of success~\cite{googleEC1,googleEC2}. Recently, the development of ``gap-engineered'' qubits has drastically lowered the logical error rate in these systems by mitigating the impact of phonon bursts in the device substrate caused by high energy ionizing particles~\cite{Sun_2012,McEwen_2024,Marchegiani_2022}. However, while these gap-engineered systems have drastically reduced correlated relaxation errors, they do not remove them entirely. Also, correlated phase errors induced by radiation impacts remain a challenge, making it critical to have the ability to remove phonons in the substrate before they break Cooper pairs~\cite{kurilovich2025correlatederrorburstsgapengineered}. We propose that coating a superconducting qubit chip with a layer of superfluid helium may drastically lower the excess quasiparticle density following radiation impacts by removing phonons before they generate quasiparticles. While superfluid-immersed qubits have been studied previously~\cite{q99s-lrnv} and show potential improvements to qubit stability, the impact of ionizing radiation in this configuration has yet to be studied. Based on our results, we believe that superconducting qubit chips may see as much as a 99\% reduction of excess phonons caused by the impacts of ionizing radiation~\cite{Iaia2022,Yelton2024,Larson2025}.

\section{Conclusion} \label{sec:conclusion}

We have demonstrated the performance of a TES-based athermal phonon calorimeter both when dry and when in contact with a superfluid $^4$He bath. We find that the energy collection efficiency of the detector is degraded by two orders of magnitude due to efficient transport of athermal phonons into the superfluid. We also find that the bias power of the detector changes noticeably when wet, implying that the superfluid either increases thermal conduction or removes a source of parasitic power. These measurements not only influence the design of future low-mass dark matter search experiments, but also inform knowledge of solid-state physics at superfluid/solid boundaries and introduce a mechanism with which to reduce phonon-induced quasiparticle poisoning in superconducting qubits.


\section{Acknowledgements}
This work was supported in part by DOE Grant DE-SC0025523 and DOE Quantum Information Science Enabled Discovery (QuantISED) for High Energy Physics (KA2401032). This material is based upon work supported by the National Science Foundation Graduate Research Fellowship under Grant No. DGE 1106400, the Department of Energy National Nuclear Security Administration through the Nuclear Science and Security Consortium under Award Number(s) DE-NA0003180 and/or DE-NA0000979, and by an appointment to the Intelligence Community Postdoctoral Research Fellowship Program at the National Institute of Standards and Technology administered by Oak Ridge Institute for Science and Education (ORISE) through an interagency agreement between the U.S. Department of Energy and the Office of the Director of National Intelligence (ODNI). This material is based upon work supported by the Department of Energy, National Nuclear Security Administration under Award Number DE-NA0004277 conducted by the Nuclear Science and Security Consortium. Work at Lawrence Berkeley National Laboratory was supported by the U.S. DOE, Office of High Energy Physics, under Contract No. DEAC02-05CH11231. Work at Argonne is supported by the U.S. DOE, Office of High Energy Physics, under Contract No. DE-AC02-06CH11357. W.G. and Y.Q. acknowledge the support by the National High Magnetic Field Laboratory at Florida State University, which is supported by the National Science Foundation Cooperative Agreement No. DMR-2128556 and the state of Florida. This work benefited from state support managed by the French National Research Agency under the France 2030 program, reference ANR-24-RRII-0001. This report was prepared as an account of work sponsored by an agency of
the United States Government. Neither the United States Government nor any agency
thereof, nor any of their employees, makes any warranty, express or implied, or assumes
any legal liability or responsibility for the accuracy, completeness, or usefulness of any
information, apparatus, product, or process disclosed, or represents that its use would
not infringe privately owned rights. Reference herein to any specific commercial product,
process, or service by trade name, trademark, manufacturer, or otherwise does not
necessarily constitute or imply its endorsement, recommendation, or favoring by the United States Government or any agency thereof. The views and opinions of authors expressed herein do not necessarily state or reflect those of the United States Government or any agency thereof.


\appendix
\clearpage

\section{Heat flow out of the Tungsten TES}
\label{appendix:Gta}

Sections~\ref{sec:methods_IV_dIdV} and~\ref{sec:discussion_IV} demonstrate that superfluid helium aids in cooling the tungsten TES, but fail to differentiate cooling via direct conduction or by a reduction in parasitic power from LEE. In this appendix, we investigate the precise nature of this cooling.

We follow the measurement prescription defined in \cite{watkinsThesis}, calculating TES bias powers at which the TES is at fixed resistance/temperature for various silicon temperatures. In the dry case, we expect the bias power $P$ (at fixed tungsten TES temperature $T_c$) and silicon temperature $T_\text{Si}$ to obey:

\begin{equation}\label{eq:Gta}
    P(T_\text{Si}) =P_0\left(1 - \left(\frac{T_\text{Si}}{T_c}\right)^\beta\right)
\end{equation}
with exponent $\beta=5$, assuming that the electron-phonon impedance in the tungsten dominates.

In the wet case, additional conductive cooling is predicted to have a small impact, as it does not circumvent the dominant electron-phonon impedance. If conductive cooling is indeed causing the increase in bias power, only the parameter $P_0$ should see a very slight change. The helium might also reduce a parasitic power caused by LEE through phonon sinking. If this is causing the increase in bias power, we would expect a change to both $P_0$ and $T_c$.

\begin{figure}
    \centering
    \includegraphics[width=\linewidth]{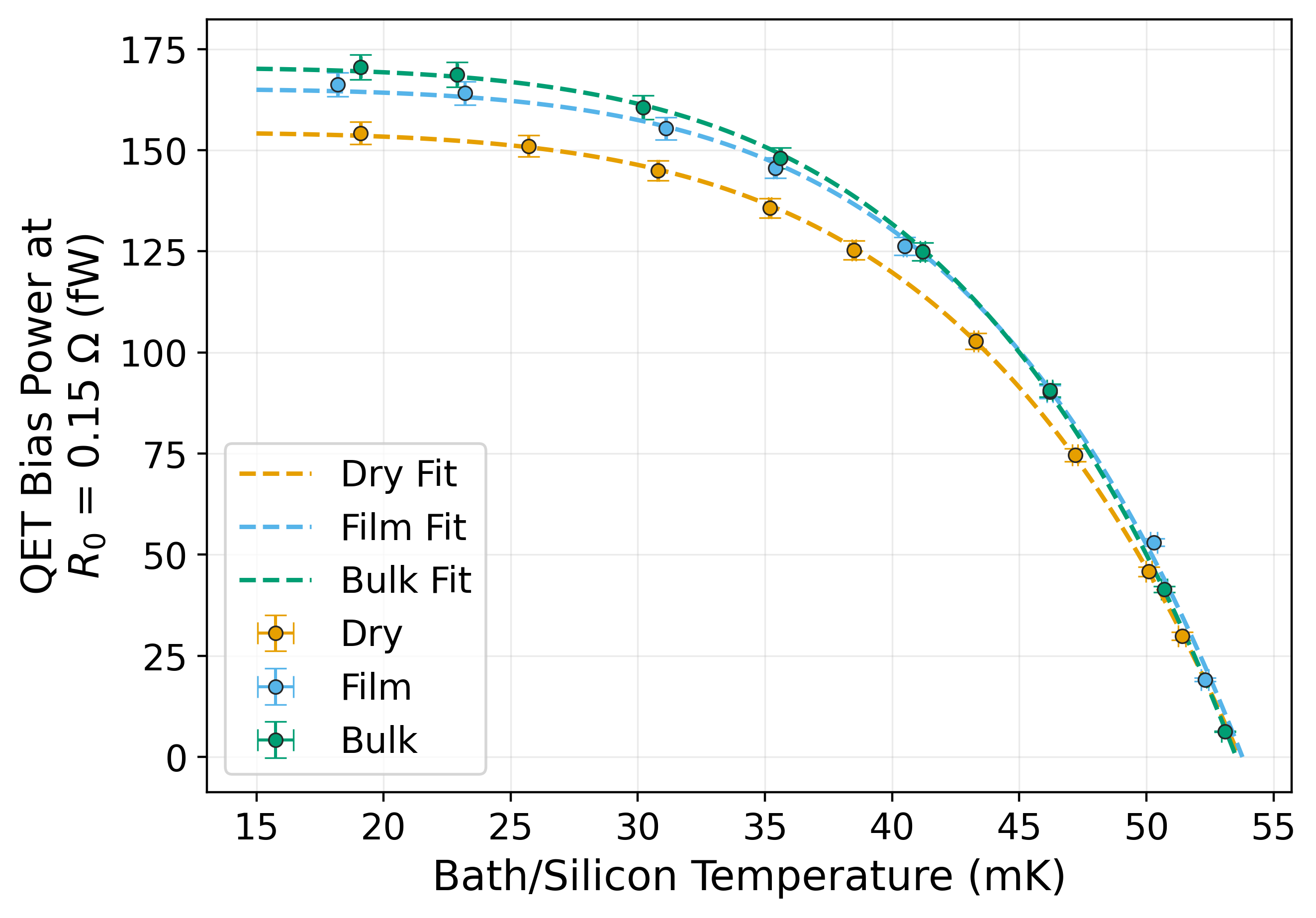}
    \caption{Bias power vs.\ silicon temperature data at fixed TES temperature for the three fill states. Fits to Eq.~\ref{eq:Gta} are superimposed as dashed lines. Fits are described in Table~\ref{tab:Gta_fit}.} 
    \label{fig:gta}
\end{figure}

\begin{table*}
    \centering
    \caption{Best-fit TES parameters for the three helium conditions.
    The first uncertainty is statistical and the second is systematic. $-2\ln\mathcal{L}$ is the maximum likelihood estimation goodness-of-fit. Systematic uncertainties on $T_c$ and $\beta$ originate from uncertainties in the temperature calibration, while those on $P_0$ are dominated by the uncertainty in the shunt resistor used to bias the PPD. They should therefore be interpreted as correlated across the three fill states.}
    \label{tab:Gta_fit}
    \begin{tabular}{l@{\qquad}c@{\qquad}c@{\qquad}c@{\qquad}c}
        \toprule
        Condition
        & $P_0$ (fW)
        & $\beta$
        & $T_c$ (mK)
        & $-2\ln\mathcal{L}$ \\
        \midrule
        Dry
        & $154.4 \pm 1.4 \pm 30.9$
        & $5.1 \pm 0.1 \pm 0.2$
        & $53.7 \pm 0.1 \pm 4.2$
        & 0.42 \\

        Film
        & $165.1 \pm 1.5 \pm 33.0$
        & $5.3 \pm 0.1 \pm 0.2$
        & $53.8 \pm 0.1 \pm 4.5$
        & 8.21 \\

        Bulk
        & $170.4 \pm 1.6 \pm 34.1$
        & $5.1 \pm 0.1 \pm 0.2$
        & $53.5 \pm 0.1 \pm 4.4$
        & 0.61 \\
        \bottomrule
    \end{tabular}
\end{table*}

Figure~\ref{fig:gta} shows the data and fits, while Table~\ref{tab:Gta_fit} summarizes these fits. As expected, the dry case is consistent with $\beta=5$. No trend in fits of $T_c$ is observed, implying that the cooling is conductive. However, the magnitude of increase in $P_0$ is larger than would be expected from excess conductive cooling alone. In other words, these observations are in tension with each other, and we are unable to make a definitive statement regarding whether the additional bias power is due to conductive cooling or a reduction in parasitic power.

\section{Heat flow out of the Silicon Substrate}
\label{appendix:Gab}

In this appendix, we describe an attempt to measure the impact of superfluid helium on thermalizing the PPD's silicon substrate. 

We follow the measurement prescription defined in~\cite{watkinsThesis}. We make use of the heater described in Sec.~\ref{sec:hardware}. In the dry case, we expect the silicon heating power $P$ (at fixed silicon temperature $T_c$) and the bath temperature $T_\text{bath}$ to obey:
\begin{equation}\label{eq:Gab}
    P(T_\text{bath}) =P_0\left(1 - \left(\frac{T_\text{bath}}{T_c}\right)^\beta\right)
\end{equation}
with $\beta=5$, assuming that the electron-phonon impedance in the gold wirebond dominates.

In the wet cases, heat generated in the heater can flow directly into the bath through the helium, making any measurement of the power put into the silicon an upper bound. We therefore do not interpret these data, but report the raw values for the interested reader.

\begin{figure}
    \centering
    \includegraphics[width=\linewidth]{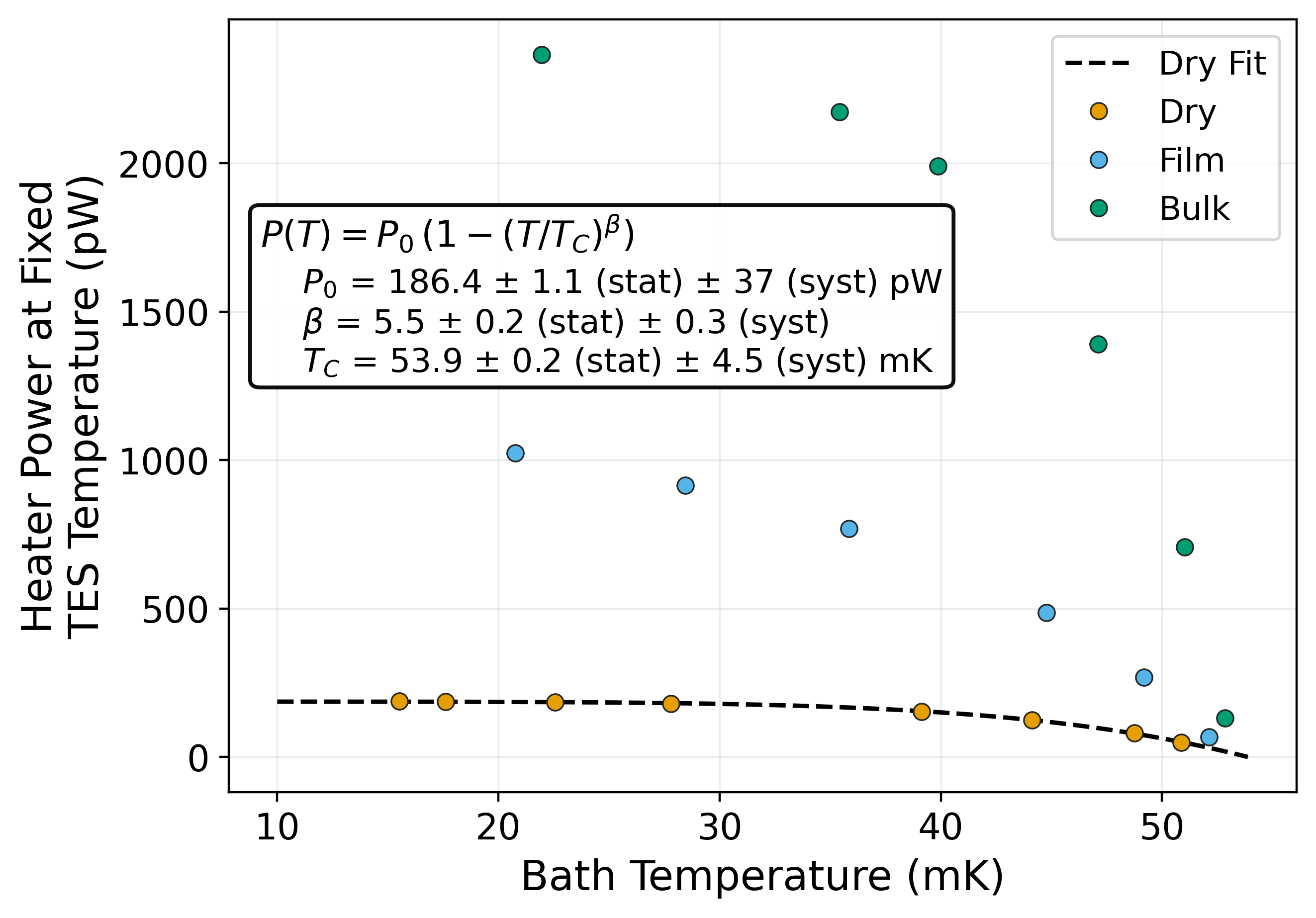}
    \caption{Heater power vs.\ bath/silicon temperature at fixed TES temperature. Fit to Eq.~\ref{eq:Gab} for the dry state is superimposed. }
    \label{fig:Gab}
\end{figure}

Figure~\ref{fig:Gab} shows the data and dry fit. The dry PPD is well-fit to Eq.~\ref{eq:Gab} and consistent with $\beta=5$.

\section{Poisson Calibration and Detector $\delta$ Parameter}
\label{appendix:poisson_cal}

Section~\ref{sec:methods_athermal} demonstrates a calibration to 3.06~eV photon absorptions in the PPD's silicon substrate. In this appendix, we attempt to calibrate using a more standard method, relying on assumptions regarding the Poisson statistics of absorbed photons. 

\begin{figure}
    \centering
    \begin{subfigure}[b]{0.45\textwidth}
        \centering
        \includegraphics[width=\textwidth]{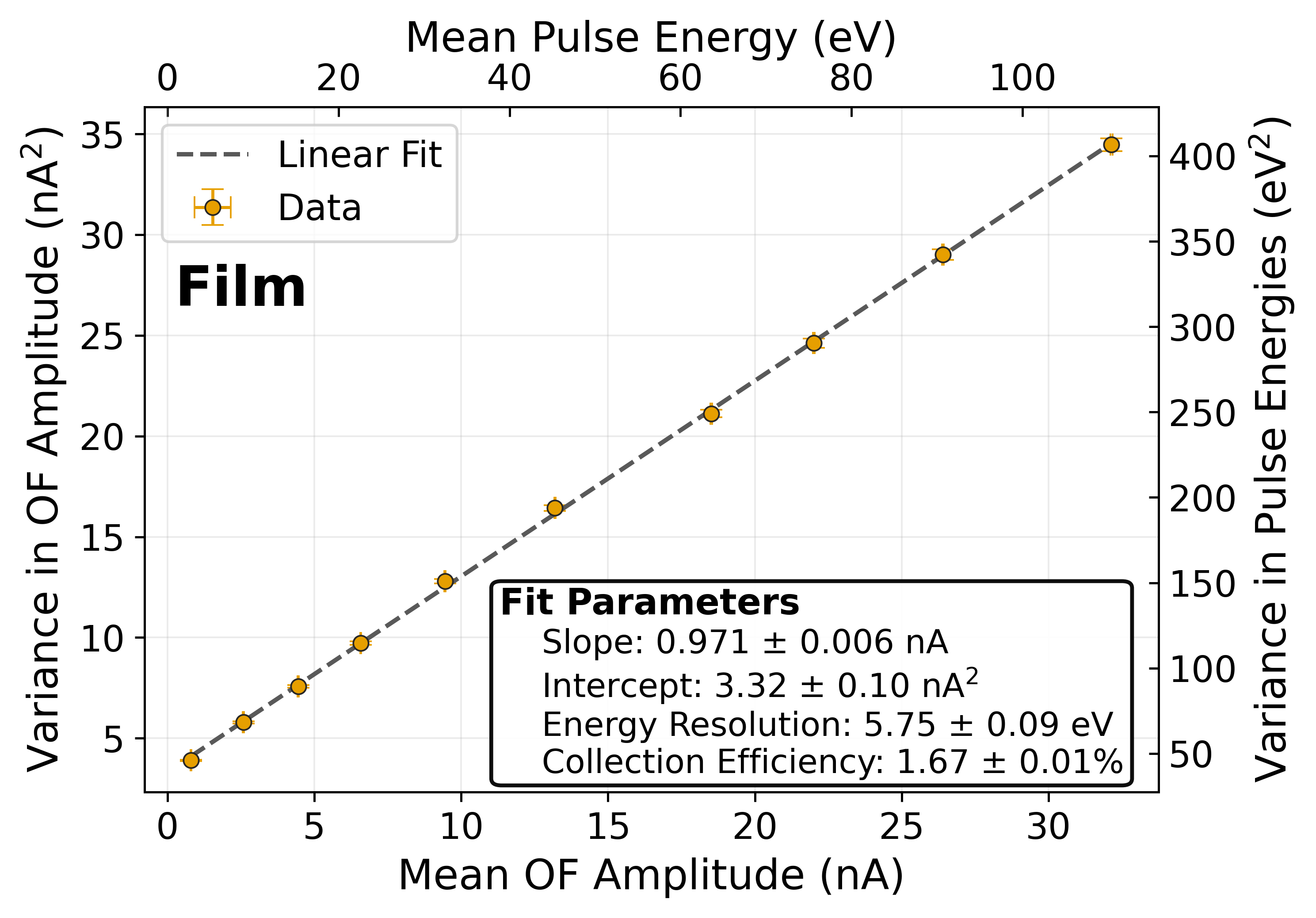}
        \label{fig:film_badcal}
    \end{subfigure}
    \hfill
    \begin{subfigure}[b]{0.45\textwidth}
        \centering
        \includegraphics[width=\textwidth]{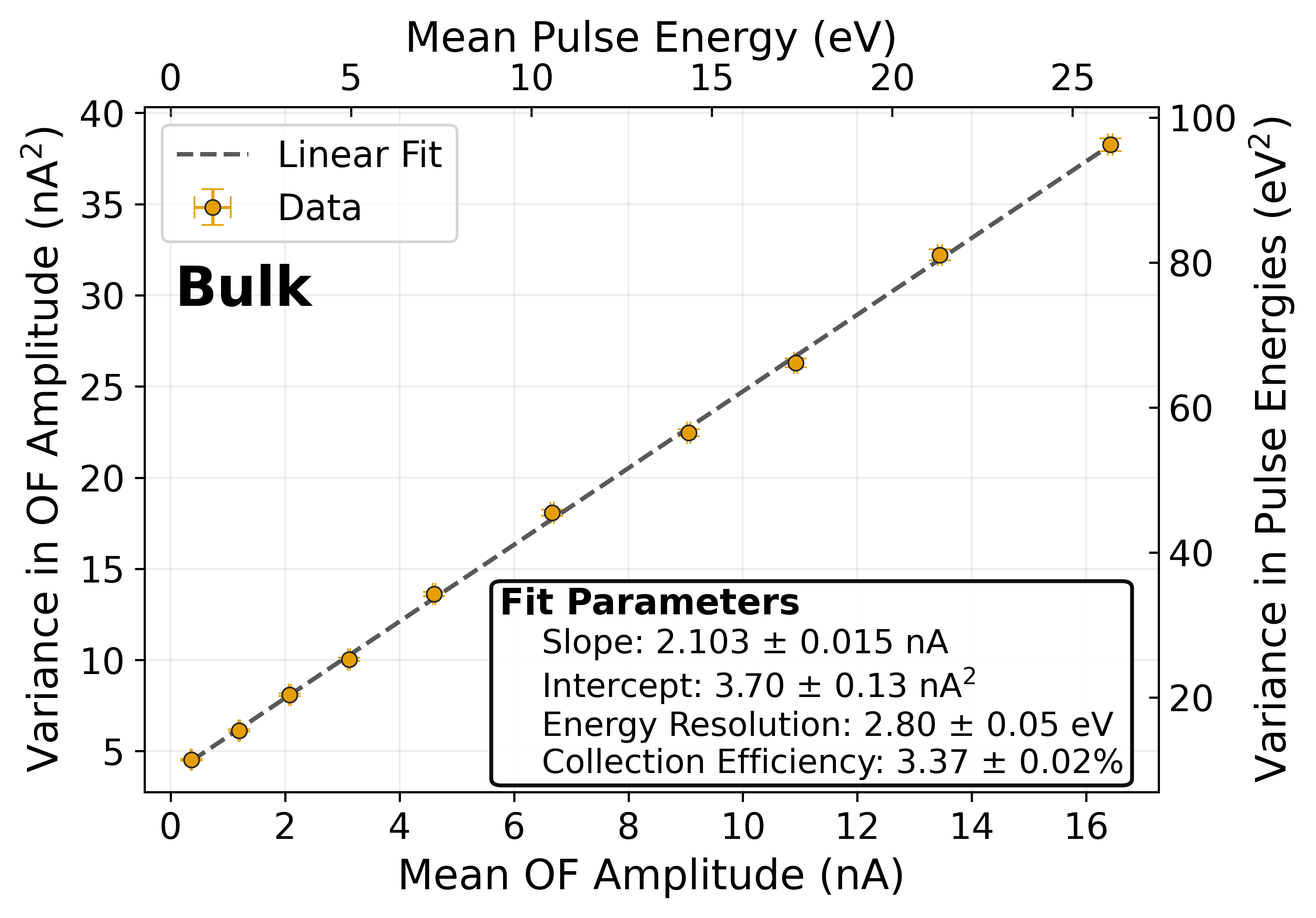}
        \label{fig:bulk_badcal}
    \end{subfigure}
    \hfill

    \caption{Mean vs. variance plots for the film (top) and bulk (bottom) states with superimposed linear fits. The implied resolutions and ECEs are reported in the inset text. These values are incorrect; see Fig.~\ref{fig:wet calibration} for the true calibration.}
    \label{fig:badcals}
\end{figure}

In each of the two wet states, photon calibration data were taken in an identical manner as described in Sec.~\ref{sec:methods_athermal} (with pulse width 1 µs and frequency 80 Hz), at a variety of laser intensities. For each of these datasets, the PPD's response to injected photons was measured offline using an OF, with trigger times set by a synchronization signal from the function generator driving the laser. The mean and variance in pulse amplitudes for each dataset were calculated and plotted in Fig.~\ref{fig:badcals}.

According to naive assumptions~\cite{Matava_calibration}, the slope of this relation defines the PPD's calibration constant/responsivity, while the intercept defines the noise due to current fluctuations. The linear fits shown in Fig.~\ref{fig:badcals} imply a resolution of $(5.75\pm0.09)$~eV in the film state and $(2.80\pm0.05)$~eV in the bulk state, in substantial tension with the results of Sec.~\ref{sec:methods_athermal}. Energy collection efficiencies are correspondingly optimistic relative to those shown in Sec.~\ref{sec:methods_athermal}.

We interpret this tension as resulting from the wet PPD's unusually high position dependence and low phonon collection, which have been argued to bias Poisson-based calibration schemes by inflating the variance of amplitude spectra beyond that expected from Poisson fluctuations alone~\cite{Matava_calibration}. 

The authors of~\cite{Matava_calibration} define a metric $\delta$ to parameterize the systematic error on such calibrations, with derived energy resolutions being too small by a factor $(1+\delta)^{-1}$. By comparing the ECEs described in Sec.~\ref{sec:methods_athermal} to those here, we measure $\delta=9.2\pm0.1$ for the film state and $\delta=20.3\pm0.2$ for the bulk state. 

While the wet PPD is unique in its poor collection and extreme position dependence, it serves as an unambiguous demonstration of the limits of Poisson-based calibration. We hope that this encourages other researchers to attempt to quantify the systematics intrinsic to this calibration method, either by means of simulation~\cite{G4CMP, G4CMP_novel} or means of steerable calibration sources that can directly measure position dependence~\cite{Stifter_MEMS, Kurinsky_MEMS}. We also note that this result also has implications for superconducting qubits, where the spatial correlation of errors is of great importance~\cite{Tan2026}. This demonstration of increased position dependence is thus another potential advantage of immersing superconducting qubits in superfluid helium.

\bibliography{apssamp}


\end{document}